\documentclass[prb,superscriptaddress,twocolumn,showkeys]{revtex4}
\usepackage{graphicx}
\usepackage{hyperref}
\usepackage{amsmath}
\usepackage{amssymb}
\usepackage{color}
\definecolor{red}{rgb}{0.8,0.,0.}

\newcommand{\BZzero}{{\rm BZ}\backslash \{0\}}
\newcommand{\ud}{\mathrm{d}}

\newcommand{\mean}[1]{\langle #1\rangle}

\newcommand{\nep}{\textrm{e}}

\newcommand{\opb}[1]{{\hat{b}^{\phantom \dagger}}_{#1}}
\newcommand{\opbdag}[1]{{\hat{b}^{\dagger}}_{#1}}
\DeclareMathOperator{\oneph}{1-ph}

\newcommand{\SL}{\scriptscriptstyle \mathrm{SL}}
\newcommand{\ext}{\mathrm{ext}}

\begin{document}

\title{Analytic understanding and control of dynamical friction}

\author{Emanuele Panizon}
\affiliation{International School for Advanced Studies, Via Bonomea 265, 34136 Trieste, Italy}

\author{Giuseppe E. Santoro}
\affiliation{International School for Advanced Studies, Via Bonomea 265,
  34136 Trieste, Italy}
\affiliation{International Center for Theoretical Physics
  (ICTP), Strada Costiera 11, I-34014 Trieste, Italy}

\author{Erio Tosatti}
\affiliation{International School for Advanced Studies, Via Bonomea 265,
  34136 Trieste, Italy}
\affiliation{International Center for Theoretical Physics
  (ICTP), Strada Costiera 11, I-34014 Trieste, Italy}
\affiliation{CNR-IOM Democritos National Simulation Center,
  Via Bonomea 265, 34136 Trieste, Italy}

\author{Gabriele Riva}
\affiliation{Dipartimento di Fisica, Universit\`a degli Studi di Milano,
  Via Celoria 16, 20133 Milano, Italy}

\author{Nicola Manini}
\affiliation{Dipartimento di Fisica, Universit\`a degli Studi di Milano,
  Via Celoria 16, 20133 Milano, Italy}


%---------------------------------------------------------
\begin{abstract}
  Recent model simulations discovered unexpected non-monotonic features
  in the wear-free dry phononic friction as a function of the sliding
  speed.
  Here we demonstrate that a rather straightforward application of
  linear-response theory, appropriate in a regime of weak
  slider-substrate interaction, predicts frictional one-phonon
  singularities which imply
  a non-trivial dependence of the dynamical friction force on the slider
  speed and/or coupling to the substrate.
  The explicit formula which we derive reproduces very accurately the
  classical atomistic simulations when available.
  By modifying the slider-substrate interaction the analytical
  understanding obtained provides a practical means to tailor and
  control the speed dependence of friction with substantial freedom.
\end{abstract}


\maketitle

%----------------------------------------------------------------------------
\section{Introduction}
%----------------------------------------------------------------------------

Friction is ubiquitous for contacting objects in relative motion.
Macroscopic mechanical energy gets converted more or less rapidly,
besides wear and other structural transformations, into thermal energy,
namely the random excitations of the vibrational degrees of freedom
(phonons) of solids, as well as electronic ones, when available.
Dissipation occurs from the largest scale of sliding geological faults
down to the nanometric and atomic scale of nanoelectromechanical devices
and atomic force microscopy (AFM) experiments.
Following the great variety of experimental and technological
approaches, the fundamentals of friction have been hystorically well
rationalized.\cite{Bowden50, PerssonBook,Robbins01, Muser06, Mate08,
  VanossiRMP13, Gnecco15, Manini15, Manini16}
Nonetheless, the corresponding theoretical efforts addressing the
underlying physics are not yet complete and satisfactory, and many
aspects still require clarification.
In particular, basic standard models introduced in the early 20$^{\rm th}$ 
century, namely the Prandtl-Tomlinson model \cite{Prandtl28,Tomlinson29} 
and the Frenkel-Kontorova model,\cite{Frenkel38, Kontorova38a, Floria96, Braunbook} 
are still largely used to date as workhorses representing phononic, wear-free static 
and sliding friction at an atomic level.
In addition, current theoretical research relies routinely on
system-specific models for molecular-dynamics (MD) simulations based on
realistic interatomic force fields.
Both the early models and the modern MD simulations usually rely on
adding dissipation to the energy-conserving atomic dynamics by means of
artificially added thermostats of some kind, most often a Langevin
thermostat.\cite{Gardiner}
Depending on the property investigated, this kind of approach may be
acceptable or problematic.
In general, since frictional dissipation transforms the mechanical work
into heat, one should worry about spurious effects brought in by the
thermostat, whose job is precisely to remove that extra heat.
After all, if the rate of frictional dissipation is what one wishes to
describe, then the results may not be unaffected by the arbitrary
downstream energy dissipation put in ``by hand'' by the thermostat.
The present work aims to provide a step forward in this understanding of
friction, by means of an analytic insight based on linear-response
theory (LRT).
Once a reliable analytic understanding of friction is obtained, even
though in a simple model, it will inevitably suggest ways to control it
and tune it with a certain freedom.

As mentioned above, the classical frictional force and power dissipation
by a slider moving on a substrate is -- in a regime where the coupling
is weak and wear and non-linear stick-slip phenomena are absent --
essentially due to the excitation of phonons.
If the slider can be considered rigid, the relevant phonon spectrum $\omega(Q)$ is that of the substrate.
The slider-substrate interaction potential, or better its Fourier
transform $V_{\ext}(Q)$, will in turn determine the strength of the
coupling generating such phonons.
In the weak interaction regime, multi-phonon processes are negligible
and the excitation of single phonons dominates the inelastic energy loss,
yielding in principle a great predictability to the resulting friction.
The frictional force $F$ will in this case be controlled by detailed
resonances which depend on the slider velocity $v_{\SL}$, on the phonon
spectrum $\omega(Q)$ and on the coupling potential $V_{\rm ext}(Q)$, giving
rise, as we will show, to a nontrivial non-monotonic behavior quite
different from the macroscopic friction laws.

%------------------------------------------------------------------------
\begin{figure}
\centering
\includegraphics*[width=0.44\textwidth,angle=0]{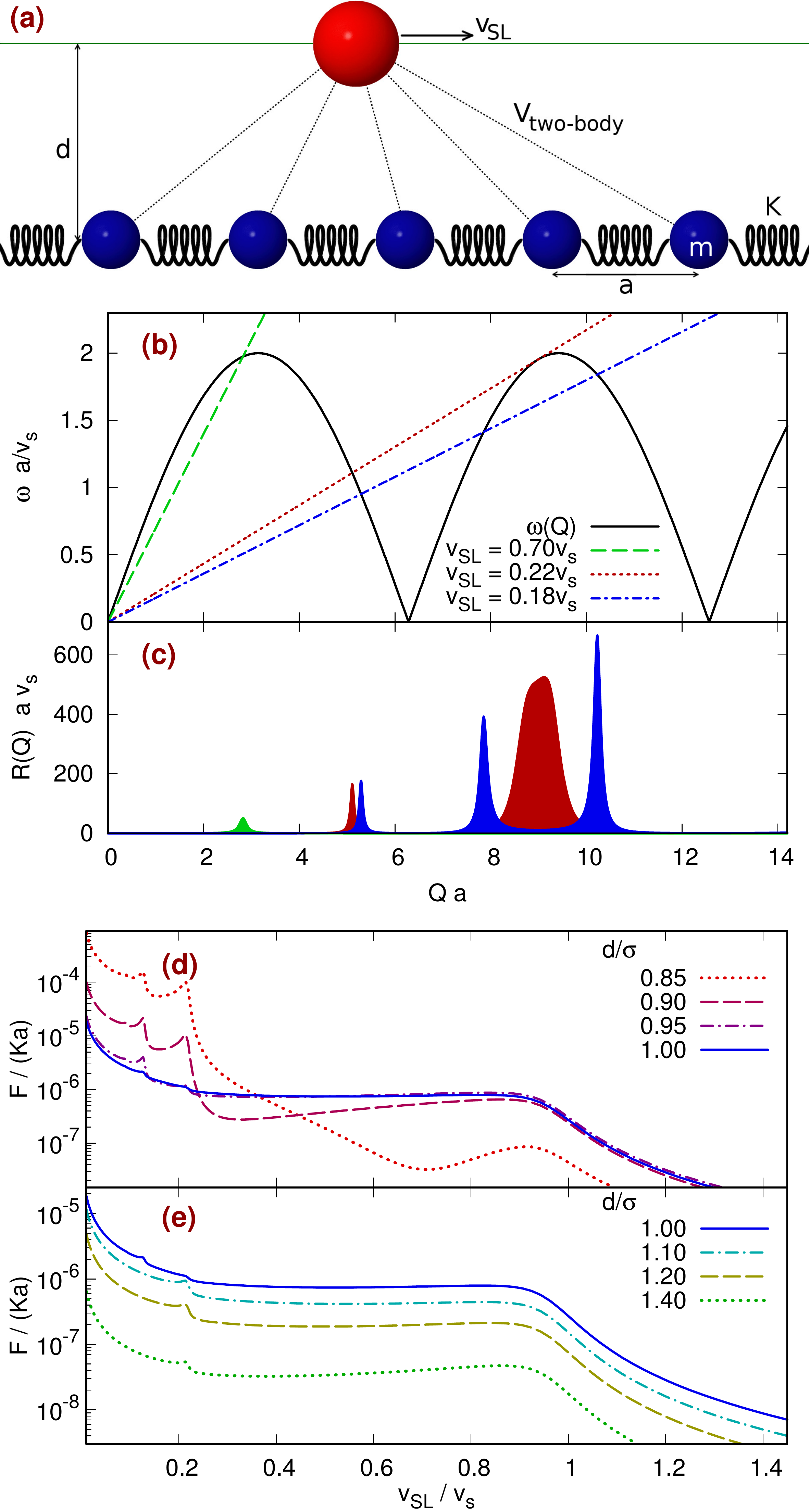} 
\caption{
  (a) A sketch of the one-dimensional model \cite{Apostoli17} considered
  in this work.
  (b) Graphical solutions of the energy-conservation condition 
  $v_{\SL}Q=\omega(Q)$ which determines the inelastically excited phonons
  for three different slider speeds $v_{\SL}$ expressed in units of
  the sound velocity $v_{\rm s}$.
  (c) The resonant magnitude evaluated as the term 
  $R(Q) = \gamma/(2\pi)\, Q^2 / \left\{[Qv_{\SL} -\omega(Q)]^2
  +(\gamma/2)^2\right\}$ of Eq.~\eqref{eq:Flangevin_intro}.
  (d,e) Friction as a function of $v_{\SL}$ at several distances $d$
  between the slider and the chain.
  $V_{\rm ext}$ is defined by a Lennard-Jones potential with $\sigma=a/2$
  and $V_0= 5\times 10^{-4}\, Ka^2$.
  The harmonic chain particle motion has a damping coefficient 
  $\gamma = 0.1 (m/K)^{1/2}$.
  Parameters $a$, $m$, and $K$ are the spacing, mass, and spring constant
  of the harmonic chain, as indicated in panel (a).
}
\label{fig:combined}
\end{figure}

In order to exemplify this physical process, we focus on the simplest model
\cite{Apostoli17} --- a point slider moving at velocity $v_{\SL}$
interacting weakly with a harmonic chain underneath, see sketch in
Fig.~\ref{fig:combined}a, through van-der-Waals forces.
Numerical simulations of this model revealed a nontrivial dependence of
friction on the sliding speed.\cite{Apostoli17}
In the following, we will develop the analytic formula for the phononic
friction in the linear-response approximation, thus providing a much more
general handle on the prediction of friction at the nanoscale.
The theory which we will set up within LRT, appropriate for the assumed
weak slider-solid interaction, is relatively simple but remarkably rich of
consequences.
The final result for the average friction force $F$ felt by the slider can
be cast in the illuminating form:
\begin{eqnarray}\label{eq:Flangevin_intro}
  F(v_{\SL}) &=&
  \frac {1}{2 m a v_{\SL}} \int_{0}^{+\infty} \!\!\! \ud Q \;  Q^2 \,  |V_{\rm ext}(Q)|^2
  \times \nonumber \\ &&
  \frac{1}{\pi}\,\frac{\gamma / 2}{[Q v_{\SL}-\omega(Q)]^2+(\gamma/2)^2} \,
   \,,
\end{eqnarray}
where $v_{\SL}$ is the slider velocity, $\omega(Q)$ the dispersion of
the sound modes of the harmonic chain and $V_{\rm ext}(Q)$ the Fourier
transform of the slider-solid interaction potential.
Here $\gamma$ is a small frictional damping constant which affects the
motion of each chain particle, representing all other degrees of freedom
coupling to the chain phonons: it results in a finite lifetime of the
phonons themselves, giving rise to a Lorentzian smearing of the resonances
that occur when the slider velocity $v_{\SL}$ matches the phase velocity
$\omega(Q)/Q$ of the phonons. 
Note that the integral over $Q$ implies an extended Brillouin Zone (BZ)
scheme for the phonons, thus not limited within $[-\pi/a,\pi/a]$.
While at very large speeds there are no solutions to the resonance condition $Q v_{\SL}=\omega(Q)$,
as $v_{\SL}$ falls below the harmonic-chain speed of sound $v_{\rm s}$ one or more solutions %of the condition  
appear, as illustrated in Fig.~\ref{fig:combined}b,c.
Sharp resonance conditions occur at critical velocities $v_{\SL}$ that match
the group velocity $v(Q)=\ud \omega/\ud Q$ at special wave-vectors $Q_i$,
an event which gives rise to a van Hove singularity in the integrand, and a
consequently sharp increase of the friction.
Remarkably, the overall shape of $F(v_{\SL})$ depends only weakly on the
small damping coefficient $\gamma$, which simply provides a smearing of the
singularities.
The crucial weighting factor in the integral is the slider-chain
interaction potential Fourier transform $|V_{\rm ext}(Q)|^2$, which in
practice could be manipulated by, e.g., modifying the slider shape and/or
its distance $d$ from the substrate.
Figure~\ref{fig:combined}d,e shows how non-monotonic and complex the
dynamical friction force $F$ can result, as a function of the slider
velocity $v_{\SL}$ and of the distance $d$ from the substrate.

This paper is organized as follows.
In Sec.~\ref{model:sec} we introduce the model, previously simulated
numerically in Ref.~\onlinecite{Apostoli17}, and recall briefly a few tools
of LRT.
We then apply them to derive the explicit decomposition of the friction
force into products of equilibrium dynamical properties of the unperturbed
chain and mechanical properties of the slider-chain interaction.
This decomposition is then used in Sec.~\ref{sec:evaluation} to evaluate
the dynamical friction force as a function of the sliding velocity for
the model at hand.
The striking parameter-free agreement of the analytical results with
previous MD simulations is presented and qualified.
The effects of the thermostat dissipation in the chain is also discussed.
In Sec.~\ref{sec:extpot} we then take advantage of the simplicity and
flexibility of the analytical result \eqref{eq:Flangevin_intro} to
investigate how changes in the slider-chain interaction potential affect
the dissipation profile, thus providing ways to tune friction, and in
particular its dependence on speed.
Conclusions are drawn in Sec.~\ref{conclusions}.

%----------------------------------------------------------------------------
\section{Model and linear-response theory} \label{model:sec}
%----------------------------------------------------------------------------

The model we consider in this work was introduced and described in
Ref.~\onlinecite{Apostoli17}.
As sketched in Fig.~\ref{fig:combined}a, it consists of a slider,
implemented in its simplest form as a point-like particle characterized
by mass $M$, position $x_{\SL}$ and velocity $v_{\SL}$,
interacting weakly via a two-body potential with each atom in a harmonic
chain characterized by particles of mass $m$, nearest-neighbor couplings
with spring constant $K$ and equilibrium spacing $a$.
The slider and the chain atoms move in one dimension (1D) along parallel
lines at a fixed distance $d$.
The slider-chain interaction energy is modelled by a sum of two-body
terms, $\sum_j V_{\rm two-body}(\sqrt{|x_j-x_{\SL}|^2 + d^2})$.
$V_{\rm two-body}$ is taken, e.g., as a (12,6) Lennard-Jones (LJ) function with
the minimum of depth $V_0$ at a separation $\sigma$.

We proceed next by assuming that the interaction between slider and chain is weak.
Under this condition, each ``collision'' occurring as the slider comes
close to particles in the chain leads to a negligible change of the
slider's momentum and kinetic energy.
As in the Born approximation of scattering theory, the slider motion is
thus conveniently approximated by an unperturbed free motion 
$x_{\SL}=v_{\SL}t$.
This is only meaningful as long as the typical interaction strength $V_0$
is much smaller than the kinetic energy of the slider $M v_{\SL}^2/2$, a
condition that can be reformulated as $v_{\SL}\gg (V_0/M)^{1/2}$.
In this approximation, the $j^{\rm th}$ particle of the harmonic chain
is influenced by a weak time-dependent external potential
$V_{\rm ext}(x_j,t)= V_{\rm two-body}(\sqrt{|x_j-v_{\SL}t|^2 + d^2})$.
We can then express the total Hamiltonian of the weakly perturbed harmonic chain as:
\begin{eqnarray} \label{Hchain:eqn}
  H_{\rm chain}(t) =
  H_{\rm harm} + \int_{-\infty}^{+\infty} \! \ud x \, V_{\rm ext}(x,t) \, n(x)   \;, 
\end{eqnarray}
where $H_{\rm harm}$ is the Hamiltonian of the unperturbed harmonic
chain.
Here $V_{\rm ext}(x,t)$ is the (small) amplitude of the perturbation,
and the density $n(x) = \sum_j \delta(x-x_j)$ is the corresponding
``operator'' to which it couples.
This formulation lends itself ideally to use LRT.\cite{Kubo:book,Vignale05}
Technically, we will adopt a quantum approach at start, which we find more
convenient: the corresponding classical LRT approach would amount to the
substitution of the quantum von-Neumann equation for the density matrix
with the corresponding classical Liouville equation.\cite{Kubo:book}
In the following we outline this calculation without omitting any useful
detail, but in lighter form -- more mathematical derivations are provided
in the Appendices.

As in Ref.~\onlinecite{Apostoli17}, we calculate the dynamical friction
force $F$ generated by sliding, which can be immediately related to the
dissipated power $W = F v_{\SL}$.
Let us focus on the {\em internal} energy $E(t)$ of the perturbed chain.
LRT tells us that:\cite{Vignale05}
\begin{eqnarray} \nonumber
  \frac{\ud}{\ud t} E(t)  &\simeq&
  - \int_{-\infty}^{+\infty} \!\! \ud x \int_{-\infty}^{+\infty} \!\! \ud x'
  \int_{-\infty}^{+\infty} \!\! \ud t' \times
  \\ \label{Et_LJ:eqn}
  && V_{\rm ext}(x,t) \, \frac{\partial \chi_{nn}^R(x,x';t-t')}{\partial t}
  \, V_{\rm ext}(x',t')
  \,,
\end{eqnarray}
where $\chi_{nn}^R(x,x';t-t')=-\frac{i}{\hbar}\theta(t-t')
\langle [\hat{n}(x,t),\hat{n}(x',t')] \rangle$
is the retarded density-density response function, the average 
$\langle \cdots \rangle$ taken on the equilibrium Gibbs ensemble, and 
$V_{\rm ext}(x,t)$ is a weak but arbitrary external potential, assumed to depend
only on $(x-v_{\SL}t)$.
The instantaneous power dissipated by the slider equals the rate of
increase of the chain internal energy $W = \ud E/\ud t$.
To evaluate the mean friction force $F$ opposing the slider motion, 
this power $W$ must be averaged over a period
$\tau=a/v_{\SL}$, which is the natural ``washboard time'' for the
slider moving across a substrate with a corrugation of period $a$:
$F=\overline{W}/v_{\SL}$.
To address this periodic problem it is advantageous to work in the
$(Q,\omega)$ Fourier domain.
In this way we can enforce the crystalline translational invariance,
leading to (see Appendix \ref{chiandF:app} for details) a simpler
expression for the average friction force in terms of the imaginary part
of the density-density susceptibility, or equivalently, via the
fluctuation-dissipation theorem,\cite{Vignale05} of the corresponding
dynamical structure factor:
\begin{eqnarray}
  F(v_{\SL}) &=& -\frac{2}{v_{\SL}} \int_{0}^{+\infty}
  \! \frac{\ud Q}{2\pi} \, \widetilde{\omega}_Q 
  \times\nonumber\\ && \qquad
  \mathrm{Im} \chi_{nn}^R(Q,Q;\widetilde{\omega}_Q) \, |V_{\rm ext}(Q)|^2 
  \nonumber \\
  \label{eq:finalW}
  &=&  \int_{0}^{+\infty} \! \frac{\ud Q}{2\pi}  \,
  Q  \left( 1 - \nep^{-\beta\hbar\widetilde{\omega}_Q} \right)  
  \times \nonumber\\ && \qquad 
  S_{nn}(Q,Q;\widetilde{\omega}_Q) \; |V_{\rm ext}(Q)|^2
  \,.
\end{eqnarray}
As required by momentum conservation in all such energy-loss problems,
in Eq.~\eqref{eq:finalW} only frequencies  $\widetilde{\omega}_Q = v_{\SL} Q$ contribute.
Equation~\eqref{eq:finalW} realizes a decomposition of the friction
force into products of the Fourier components of the structure factor of
the unperturbed chain at thermodynamic equilibrium, and of squared
Fourier components of the slider-chain interaction potential.
On one hand, this kind of decomposition is quite standard in all
applications of the LRT, as in the ordinary Born approximation of
scattering theory.
Application of this method to a proper sliding-friction problem was rather
rarely attempted before (see however Refs.~\onlinecite{PerssonTosatti99,
  Kisiel15}).
The present application demonstrating an exceptionally high accuracy of the
results represents a major milestone of the present paper.
The reader must be warned that the strict applicability of LRT is limited
to smooth-sliding regimes where the system remains clear of highly
nonlinear effects such as stick-slip dynamics or wear, which can and do
occur in the physics of friction.\cite{Granato04, Urbakh04, Santoro06,
  Manini08Erice, Manini15, Manini16}
In the present weak-perturbation approach however, the Prandtl-Tomlinson
smooth-sliding condition
\footnote{
  The Prandtl-Tomlinson condition $4 \pi^2 U_0/Ka^2 \ll 1$ involves the
  ratio between the substrate corrugation energy $U_0$ and the typical
  elastic energy stored in the driving spring of stiffness $K$ when
  elongated over a corrugation lattice spacing $a$.\cite{Vanossi13}
}
is always automatically satisfied.
For smooth sliding therefore, one can take advantage of the analytical
predictive power of the LRT decomposition, as we shall illustrate in the
following sections.

%----------------------------------------------------------------------------
\section{Evaluation of the dynamical friction force}\label{sec:evaluation}
%----------------------------------------------------------------------------

\begin{figure}
\centering
\includegraphics*[width=0.5\textwidth,angle=0]{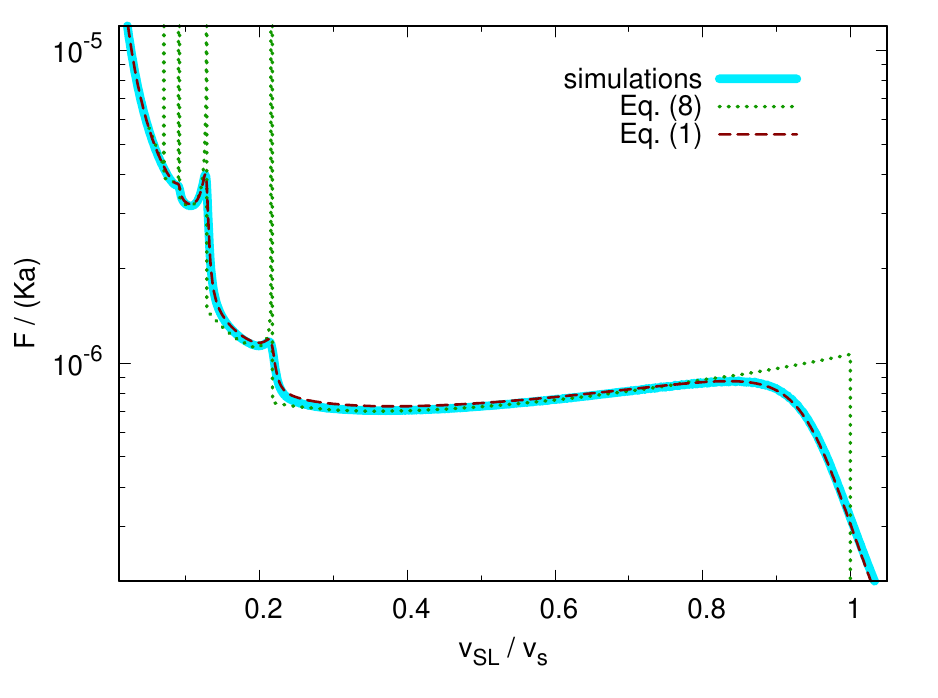}
\caption{
  Comparison of the slider-speed dependence of the friction force $F$
  obtained from the two analytical expressions \eqref{eq:attrito}
  (dotted, no dissipation included) and \eqref{eq:Flangevin_intro} (dashed,
  dissipation included), with that obtained for same conditions by
  numerical simulations carried out for a chain of 500 atoms
  (solid).\cite{Apostoli17}
  All calculations have spacing $a$, nearest-neighbor spring constant
  $K$, damping rate $\gamma=0.1 \, (m/K)^{1/2}$, LJ slider chain
  interaction with $\sigma=0.5 \, a$, and distance $d=0.475 \, a$.
  Peaks for $v_{\SL} \leq 0.22 v_{\rm s}$ correspond to speeds listed
  in Table~\ref{tbl:sol}, and evolve from true divergences without
  dissipation to sharp peaks with dissipation.
  The agreement between the analytical and the simulation results is
  genuinely striking.
}
\label{fig:compare}
\end{figure}

We now apply the general LRT prescription in Eq.~\eqref{eq:finalW}
--- which with minor adjustments would describe more realistic
slider-substrate situations in higher dimensions --- to our toy problem
of a 1D harmonic chain substrate perturbed by a point
slider a distance $d$ away.
The exact expression for the harmonic-chain structure factor
$S_{nn}(Q,Q,\omega)$ --- see Appendix \ref{structure:sec} for a
derivation --- is reported in Eq.~\eqref{eq:snn}, but is practically
impossible to use.
To proceed, we make use of a relatively standard {\em one-phonon approximation},\cite{Ashcroft} 
which is quite reasonable in higher dimensions, ignoring here all pathologies typical of 1D, due in turn to
singularities generated by the $1/\omega(Q)$ factor appearing in the
$Q$-integrals.
Taken literally --- see Appendix \ref{structure:sec} for a discussion of
the relevant steps and subtleties --- the one-phonon approximation leads to
the following expression for the dynamical structure factor:
\begin{equation} \label{eq:dyn_S_1ph_main}
  S_{nn}^{\oneph}(Q,Q;\omega>0) =\frac{\pi Q^2 }{m a \omega(Q)}  \,
  \frac{1}{1-\nep^{-\beta\hbar\omega(Q)}} \, \delta(\omega-\omega(Q))
\,,
\end{equation}
where $\omega(Q)$ is the phonon dispersion in the {\em extended} BZ
scheme.
Substituting this approximate structure factor in the expression for the
friction force, Eq.~\eqref{eq:finalW}, we obtain:
\begin{widetext}
\begin{equation}
F(v_{\SL}) =  \int_{0}^{+\infty} \! \frac{\ud Q}{2\pi} \; Q
  \left(1 - \nep^{-\beta\hbar v_{\SL} Q} \right) \, 
  \frac{\pi Q^2 }{m a \omega(Q)} \;
  \frac{1}{1-\nep^{-\beta\hbar\omega(Q)}}
  \; \delta(v_{\SL} Q-\omega(Q)) \; |V_{\rm ext}(Q)|^2  \;.
\end{equation}
\end{widetext}
The integral over $Q$ is now easy to perform using the well-known
property of the Dirac delta:
\begin{equation}
  \int_{-\infty}^{\infty} \ud Q \; g(Q) \; \delta(f(Q)) = \sum_i\frac{g(Q_i)}{|f'(Q_i)|}\;,
\end{equation}
where $Q_i$ are the solutions of $f(Q)=0$.
In the present case, $f(Q)=v_{\SL} Q-\omega(Q)$ and $f'(Q)=v_{\rm
  SL}-v(Q)$, where $v(Q)= \ud \omega /\ud Q$ is the group velocity of
the phonon dispersion.
Substituting and simplifying, we conclude that:
\begin{equation}\label{eq:attrito}
  F(v_{\SL}) = \frac {1}{2 m a v_{\SL}}
  \sum_{i} \frac{ Q_i^2} {|v_{\SL}-v(Q_i)|} |V_{\rm ext}(Q_i)|^2
  \,.
\end{equation}
Here the values $Q_i$ are the solutions of the momentum-conservation
equation
\begin{equation}\label{eq:reson}
  v_{\SL} Q = \omega(Q) \,.
\end{equation}
Note that both $\hbar$ and the inverse temperature $\beta$, which
still enter the expression for $S_{nn}^{\oneph}(Q,Q;\omega>0)$, have
remarkably disappeared from the final expression for $F(v_{\SL})$.
The expression \eqref{eq:attrito} for $F(v_{\SL})$ compares in
surprisingly accurate detail against earlier {\em classical} numerical
simulations for the 1D toy problem,\cite{Apostoli17} see Fig. \ref{fig:compare}.
Indeed all the important frictional features --- the overall non-resonant
friction, and the positions and relative strengths of the resonant peaks
--- are reproduced.
The overall quantitative agreement is impressive, except for the
divergences corresponding to the vanishing denominators in
Eq.~\eqref{eq:attrito}, which in the simulation are replaced by rounded
peaks.
%------------------------------------------------------------------------
%------------------------------------------------------------------------
%
%

The reason for the smoothing of singularities can be attributed to the
presence of a (small but finite) viscous force $-\gamma \dot x_j$
introduced in the classical numerical simulations \cite{Apostoli17} in
order to dispose of the phonon energy generated by the slider before they
return to the contact point through the boundary conditions.
We expect that {\em in general} the unavoidable presence of dissipation and
anharmonicity in the substrate will lead to a decay of the density-density
correlation function for large $t$:
\begin{equation}\label{eq:Slangevin}
  S_{nn}^{\rm diss}(x,x';t) = S_{nn}(x,x';t)
  \; \nep^{- \frac{\gamma}{2} \, |t| }  \,.
\end{equation}
This decay will in turn lead to a {\em broadening} of the
$\delta(\omega-\omega(Q))$ appearing in the one-phonon structure factor in
Eq.~\eqref{eq:dyn_S_1ph_main}:
\begin{equation}
\delta\left(\omega-\omega(Q)\right) \to \frac{1}{\pi}
\frac{\gamma/2}{(\omega-\omega(Q))^2 + (\gamma/2)^2 } \,.
\end{equation}
Hence, accounting for dissipation one obtains the friction-force
expression \eqref{eq:Flangevin_intro}.
This final expression is possibly even more straightforward for a numerical
evaluation than Eq.~\eqref{eq:attrito}.
As shown by the dashed line in Fig.~\ref{fig:compare}, the resulting
friction reproduces {\em quantitatively} the simulated results at all
speeds with no fitting parameter, taking for the damping rate the same
value $\gamma = 0.1 \, (m/K)^{1/2}$ used in the
simulations.\cite{Apostoli17}

We now discuss the physical insights present in the expressions 
Eq.\eqref{eq:Flangevin_intro} and  Eq.~\eqref{eq:attrito}. 
Let us focus, for clarity of presentation, on the latter, which is slightly more transparent in notation, 
while qualitatively similar to the former.
We observe first that at any given speed $v_{\SL}$ all contributions to friction %(not just the peaks)
come from a small set of phonon modes which are the solutions of the
relation \eqref{eq:reson}, $v_{\SL} Q=\omega(Q)$, equating the slider speed to the phase
velocity of the phonons.\cite{Apostoli17}
For nearest-neighbor springs the chain dispersion relation is
$\omega(Q)=2 v_{\rm s} a^{-1} \, |\sin (Qa/2)|$.
The condition $v_{\SL} Q=\omega(Q)$ is thus conveniently 
rewritten in dimensionless form
\begin{equation}\label{eq:reson_dimensionless} 
\bar{v}_{\SL} \bar Q = 2 \left| \sin\left({\bar Q}/{2}\right) \right| \,,
\end{equation}
where $\bar Q=Qa$ and $\bar{v}_{\SL}=v_{\SL}/v_{\rm s}$ is the
ratio between the slider speed and the speed of sound 
$v_{\rm s} = a (K/m)^{1/2}$, and we introduce the dimensionless dispersion
$\bar\omega(\bar Q) = a v_{\rm s}^{-1}\, \omega(\bar Q/a) = 2 |\sin(\bar Q/2)|$.
% NICK: nella vers. prima questa equazione era scritta tutta toppata!
% Ora mi pare corretta.

For a given (dimensionless) slider speed $\bar{v}_{\SL}$, each intersection of
the straight line $\bar{v}_{\SL} \bar Q$ with the phonon dispersion
$\bar\omega(\bar Q)$ determines, regardless of details of the
slider-chain interaction, a contribution to friction.
This is illustrated in Fig.~\ref{fig:combined}b for three velocities.
For $\bar{v}_{\SL}>1$, the resonant phonon intersections cease, and
Eq.~\eqref{eq:attrito} predicts frictionless sliding, as reflected by the
sharp drop of the dotted line in Fig.~\ref{fig:compare}.
As $\bar{v}_{\SL}$ decreases below unity, initially
Eq.~\eqref{eq:reson} has a single solution $\bar Q_1$ in the $[0, 2 \pi)$
% NICK: corretto qui, c'era scritto $[0, \pi)$, ma era sbagliato! See fig.
interval, as for the dashed line in Fig.~\ref{fig:combined}b.
Then, starting from $\bar{v}_{\SL} \lesssim 0.217$, two new solutions
appear at $2\pi<\bar Q_2\leq \bar Q_3<4\pi$,
% NICK: c'era scritto $\pi<\bar Q_2\leq \bar Q_3<3\pi$, ma era sbagliato!
as for the dotted and dot-dashed lines in Fig.~\ref{fig:combined}b.
More solutions appear in pairs at larger $\bar Q$ as $\bar{v}_{\SL}$
is further reduced --- therefore friction grows.

According to Eq.~\eqref{eq:attrito}, friction is determined by the
magnitude of the squared Fourier transform of the external potential at the
resonating wave vectors $Q_i=\bar Q_i/a$.
These magnitudes are summed with weights given by $Q_i^2/|v_{\SL}-v(Q_i)|$. 
This leads to the emergence of sharp resonance conditions when $v(Q_i) \sim v_{\SL}$, similar to those reported earlier in the
Frenkel-Kontorova model.\cite{Strunz98a,vandenEnde12}
The resulting divergences in $F$ resemble van Hove singularities \cite{vanhove53} in 1D.
The regular van Hove singularities, e.g.\ in the density of states, are
associated to regions of the BZ where the group velocity vanishes, i.e.
$\ud\omega(Q) /\ud Q=0$.
In the present case the singularities are generated by the condition $Q v_{\SL}=\omega(Q)$, 
and lead to {\em divergent friction when the slider velocity, the phase velocity, and the group velocity coincide}:
\begin{equation} \label{eq:peaks}
  {v}_{\SL} = \frac{\omega(Q)}{Q}  
                 = \frac{\ud\omega(Q)}{ \ud Q}  \equiv v(Q) \,.
\end{equation}
Singularities of the same origin are long known in energy-loss problems,
such as for example in the crossing of solids by fast electrons,\cite{Lindhard54} 
or by neutrons,\cite{AshcroftAppN} or even in inelastic helium scattering at solid surfaces.\cite{Cabrera69}

Here, the resonances have a physically intuitive explanation.
The slider moving at speed $v_{\SL}$ can only excite those phonons
with matching phase velocity $\omega(Q)/Q$ --- their ``wave crest''
velocity matching that of the slider.
However, only when the exciting particle and the excited phonon 
wave-packet, moving with the group velocity $v(Q)$,
``fly'' together for a long time the energy transfer between them is
really strong.
Thus the effectiveness of the excited phonons to download and carry
energy away is strong --- resonant --- when all velocities coincide, and
weaker when they do not match.
Crudely speaking, only at resonance the slider ``surfs'' the phonon wave
crest appropriate for its speed.

%------------------------------------------------------------------------
\begin{table}
  \begin{center}
    \begin{tabular}{c|c|c|c} \hline \hline 
      $j$&$\bar{v}_{{\SL}\,j}$&$\bar Q_j$ & $\bar Q_j$ (1$^{\rm st}$ BZ)\\
      \hline  \hline
      1& $1$		&$0$	   &$0$ 	 \\ \hline
      2& $0.217243$&$8.986819$&$2.70364$\\ \hline
      3& $0.128375$&$15.45050$&$2.88413$\\ \hline
      4& $0.091325$&$21.80824$&$2.95869$\\ \hline
      5& $0.070914$&$28.13239$&$2.99965$\\ \hline
      6& $0.057972$&$34.44151$&$3.02558$\\ \hline
      7& $0.049030$&$40.74261$&$3.04349$\\ \hline
      8& $0.042480$&$47.03890$&$3.05661$\\ \hline
      9& $0.0374745$&$53.33211$&$3.06663$\\ \hline
     10& $0.0335251$&$59.62320$&$3.07453$\\ \hline \hline           
    \end{tabular}
  \end{center}
  \caption{\label{tbl:sol}
    Resonant dimensionless velocities $\bar{v}_{{\SL}\,j}$, and
    corresponding solutions $\bar Q_j=Q_ja$ for the first ten friction peaks for
    decreasing $\bar{v}_{\SL}=v_{\SL}/v_{\rm s}$, evaluated from the tangency
    condition, Eq.~\eqref{eq:peaks}.
    The last column reports the values of $\bar Q_j$ %(also dimensionless)  
    folded back to the first BZ  $(-\pi, \pi]$.
  }
\end{table}	
%------------------------------------------------------------------------

Geometrically the divergences predicted occur when the 
$v_{\SL} Q$ straight line is tangent to the dispersion curve
$\omega(Q)$ (such as the dotted line of Fig.~\ref{fig:combined}b): each divergence
corresponds to a solution of Eq.~\eqref{eq:peaks}, 
which marks precisely the appearance (disappearance) of a new pair of
solutions of $v_{\SL} Q=\omega(Q)$ as $v_{\SL}$ is decreased
(increased).
Table \ref{tbl:sol} reports the numerical values of the ten largest
resonant speeds at which such divergences appear.
We stress that these resonant speeds are uniquely functions of the chain
dispersion relation, therefore of its structure factor, independent of
the ``form factor'', namely the slider-chain interaction $V_{\rm ext}$.
As for the slider-chain interaction, the theory poses no significant
restriction on the shape of the weak potential $V_{\rm ext}$ beyond that
of a a dependence on time and space of the form $(x-v_{\SL} t)$,
i.e.\ $V_{\rm ext}$ must have a fixed profile translating rigidly at a
speed $v_{\SL}$, and it must be possible to evaluate its Fourier
transform.

%----------------------------------------------------------------------------
\section{Varying the Slider-Chain Interaction}\label{sec:extpot}
%----------------------------------------------------------------------------

As shown above, Eq.~\eqref{eq:Flangevin_intro} compares extremely well
with MD simulations carried out with a LJ slider-chain potential.
There is no reason to believe it would not provide equally reliable
friction evaluations for other physically meaningful external potentials.
In the following we study the effect on friction of changes in
$V_{\rm ext}$.

%------------------------------------------------------------------------
\begin{figure}
\centering
\includegraphics*[width=0.5\textwidth,angle=0]{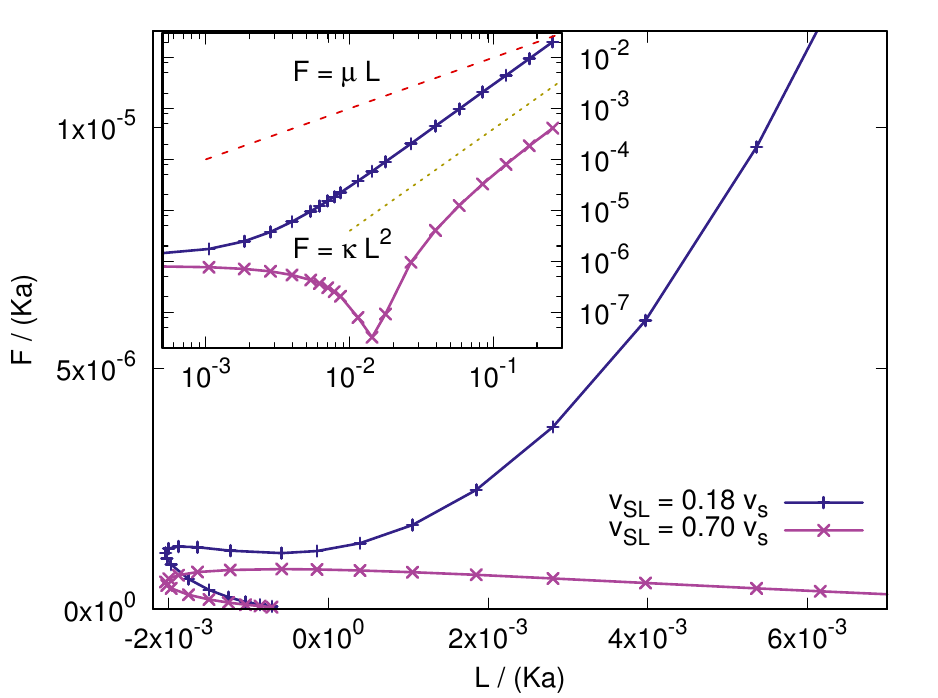}
\caption{\label{fig:amontons}
Load dependence of the dynamic friction force computed according to
Eq.~\eqref{eq:Flangevin_intro} for a large range of distances at two
sliding speeds.
Comparison with a linear increase (Amonton-like) is also shown.
The effective load $L$ is estimated as the average vertical force
acting on the slider in one sliding period, Eq.~\eqref{eq:load}.
The model parameters are the same as for Fig.~\ref{fig:combined}.
The connecting lines are guides to the eye.
In the inset the scale is logarithmic on both axes to better illustrate
the quadratic scaling of friction for large loads, at all speeds.
At small positive loads, friction becomes very nonmonotonic, and even
exhibits a deep minimum at the larger considered speed.
The reentrance of both friction curves at negative loads exhibits the
standard attractive, large-distance regime observed in AFM experiments.
}
\end{figure}
%------------------------------------------------------------------------

We first stick to the basic LJ potential form, and investigate the
effect of varying the distance $d$ of the slider from the surface, which
is somewhat equivalent to varying the load applied in an AFM experiment.
Some interesting novelties that emerge are shown in
Fig.~\ref{fig:combined}d,e.
First of all, as expected, regardless of $d$ the resonant friction peaks
occur at the same speeds, listed in Table~\ref{tbl:sol}, determined
purely by the chain dispersion relation.
By contrast, the absolute and relative dissipation at the resonances, and
in between, do change when $V_{\rm ext}$ is varied.
We observe in particular a gradual weakening of the low-speed peaks as
$d$ is increased.
The peaks emerge from solutions of Eq.~\eqref{eq:peaks} at large
wavevectors $Q$, that correspond to potential variations at very short
length-scale.
As the distance is increased the interaction between slider and substrate
smoothens out, large-$Q$ Fourier components decrese dramatically, and
therefore only small wavevectors contribute to the dissipation.
As a result, at low speed $\bar{v}_{\SL}<0.22$ friction grows
monotonically as the perturbation strengthens at shorter distances.
By contrast, at larger speed friction shows a nonmonotonic behavior as a
function of $d$.
Since sliding assumes in our model a fixed distance between slider and
the substrate line, the vertical (i.e.\ perpendicular to sliding) force
component oscillates in time.
For each distance $d$, we can introduce an ``average'' load $L$, as the
vertical force experienced by the slider averaged over one period
$a/v_{\SL}$, namely
\begin{eqnarray}\label{eq:load}
  L&=&
  -\frac{v_{\SL}}a \int_0^{a/v_{\SL}}\!\!\ud t \;
  \sum_j \frac {\ud V_{\rm ext}(j a,t)}{\ud d}
  \nonumber\\&=&
    -\frac 1a \int_{-\infty}^\infty \!\ud x \;
     \frac {\ud V_{\rm ext}(x,0)}{\ud d}
   \,.
\end{eqnarray}
$L$ is positive in the repulsive region $d\lesssim \sigma$, and turns
negative in the attractive region $d \geq \sigma$.
With this definition, we can construct the friction-load curve by
varying the distance $d$ for any given value of the slider speed.
Figure~\ref{fig:amontons} reports two such curves for $\bar{v}_{\rm
  SL}=0.7$ and $\bar{v}_{\SL}=0.18$.
At low speed and for sufficiently large load (slider close to the
substrate) a joint increase of friction $F$ and load $L$ is observed.
The friction increase with load is to an excellent approximation
represented by $F \propto L^2$ rather than linear with $L$ as in, e.g.,
Amonton's law of macroscopic friction.
The reason for this quadratic friction increase with load is understood
as connected to the inverse power-law repulsive behavior ($V_{\rm ext}
\propto d^{-12}$ of the LJ interaction, but the same would hold for
another exponent) at short range.
In this regime indeed
$L \propto V_{\rm
  ext}$, whence $F \propto |V_{\rm ext}|^2$ --
Eq.~\eqref{eq:Flangevin_intro} -- entails $F \propto |L|^2$.

The additional novelty at large speed ($\gtrsim 30\%$ of the sound
velocity) is nonmonotonicity of friction versus load.
In this regime, friction initially decreases for increasing load (as
also noted in Fig.~\ref{fig:combined}d) until it nearly vanishes and
then increases again.
The reason for this frictional dip -- which would be a zero in the
$\gamma\to 0$ limit -- is a change of sign of the Fourier-transformed
interaction $V_{\rm ext}(Q)$ caused in turn by real-space
attractive-repulsive-attractive oscillations which occur along $x$ when
the load is increased in a certain range.
This kind of matrix-element zero is the straight analog of the ``Cooper
zero'' well-known in atomic spectra~\cite{FanoCooper68}.
In our case, at any velocity, the dominant contributions to friction
come from the wavevectors given by Eq.~\eqref{eq:reson} (and their
surroundings of order $\gamma / v_{\rm s}$), which for $\bar{v}_{\SL}
= 0.7$ correspond to one single region around $Q \simeq 2.8204/a$, see
Fig.~\ref{fig:combined}b,c.
For $d > \sigma$, $|V_{\rm ext}(Q)|$ has a single peak for $Q = 0$.
However for $d \gtrsim \sigma$ the interaction develops a small
repulsive contribution when the slider is atop a particle of the
substrate, generating a second peak in $|V_{\rm ext}(Q)|$ at some finite
$Q > 0$.
As $d$ drops  this second peak becomes more and more prominent, and en
route there is a wavelength where $|V_{\rm ext}(Q)| = 0$.
For $\bar{v}_{\SL} = 0.7$ this anti-resonance condition overlaps the
resonant wavevector $Q \simeq 2.8204/a$ for the value of distance $d
\simeq 0.43\, a$.
This corresponds exactly to the friction dip.
For $\bar{v}_{\SL} = 0.18$, Eq.~\eqref{eq:reson} yields three
resonant $Q$s, so that the vanishing of $V_{\rm ext}(Q)$ suppresses only
one of the three, with a much weaker overall effect.

%------------------------------------------------------------------------
\begin{figure}
\centering
\includegraphics*[width=0.5\textwidth,angle=0]{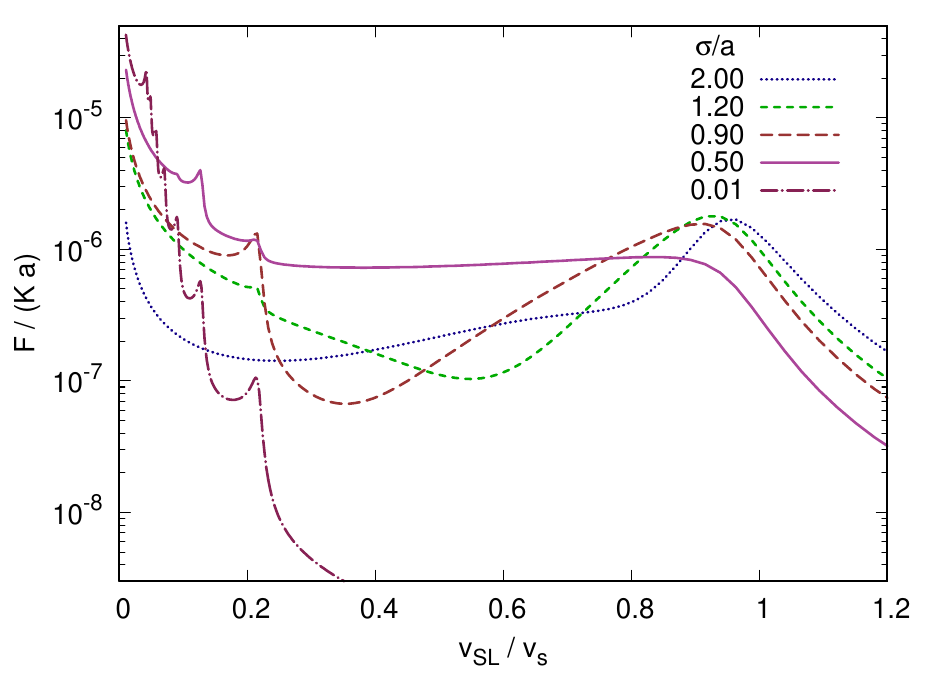}
\caption{\label{fig:sigma}
Friction-velocity profiles for obtained for varying $\sigma$ in the LJ
slider-chain interaction potential.
Distance is always $d = 0.92 \, \sigma$.
Other parameters are slider interaction depth $V_0 = 5\times 10^{-4}
Ka^2$ and damping rate $\gamma = 0.1 \, (m/K)^{1/2}$.
}
\end{figure}
%------------------------------------------------------------------------

Variations of the LJ parameter $\sigma$ characterizing the range of
action of the slider-chain interaction also affect friction through
changes in the Fourier transform $V_{\rm ext}(Q)$.
To investigate this effect we keep the distance at a fixed fraction
$d=0.92 \, \sigma$, corresponding to approximately vanishing load.
Whenever $\sigma$ is significantly larger than the substrate lattice
parameter $a$, the interaction spreads over a large number of harmonic
beads, with the result that the Fourier transform of the LJ potential is
non-negligible only for small wavevectors.
As a result, as shown in Fig.~\ref{fig:sigma}, for larger $\sigma$ the
only prominent resonant feature is the one near the speed of sound.
By contrast if $\sigma \simeq a$ several Fourier components contribute
similarly to $V_{\rm ext}(Q)$, resulting in several visible frictional
features at the resonant speeds of Table~\ref{tbl:sol}.
In the limit $\sigma \ll a$ the slider interacts and collides with the
chain particles one at a time, resulting in a sharply position-dependent
potential, characterized by large high-$Q$ Fourier components.
As a result, dissipation picks up robust contributions from large
wavevectors, resulting in many resonances visible at low speed
(dot-dashed curve in Fig.~\ref{fig:sigma}).

Next, we checked the effect a different slider-chain potential form.
Comparing a Morse potential characterized by the same equilibrium
distance $\sigma$, potential well depth $V_0$, and curvature at the
minimum as the LJ potential, the resulting friction-velocity curves (not
shown) are qualitatively and quantitatively quite similar to those of
Fig.~\ref{fig:sigma} for $d$ of the order of $\sigma$.

%------------------------------------------------------------------------
\begin{figure}
\centering
\includegraphics*[width=0.5\textwidth,angle=0]{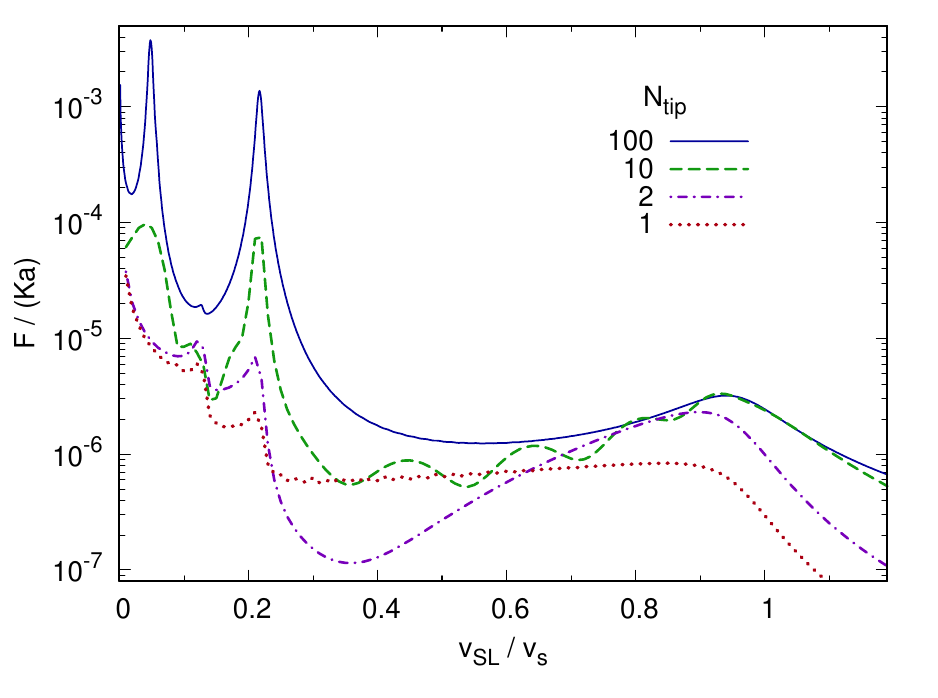}
\caption{\label{fig:tips}
  Friction profile for simulated composite tips consisting in the sum of
  $N_{\rm tip}$ equal LJ contributions displaced at intervals of $l=0.7 a$.
  All other parameters are the same used previously: $\sigma=0.5 \, a$,
  $d = 0.465 \, a$, $V_0 = 5\times 10^{-4} Ka^2$, and $\gamma = 0.1 \,
  (m/K)^{1/2}$.
}
\end{figure}
%------------------------------------------------------------------------

More interestingly, we can exploit the simple structure of the analytic
result of Eq.~\eqref{eq:Flangevin_intro} to investigate a basic model
for an extended tip.
We construct one by building the external potential as the sum of
$N_{\rm tip}$ LJ potentials, placed at fixed separation $l$ one from the
next:
\begin{equation}
  V_{\rm ext}=
  \sum_{i=1}^{N_{\rm tip}}
  V_{\rm LJ}\!\left(\left[(x-il-vt)^2+d^2\right]^{1/2}\right) \;.
\end{equation}
For $N_{\rm tip}=1$ we recover the point-like slider discussed previously.
In the limit $N_{\rm tip}\to\infty$ the slider generates a periodic
potential, so that the model should resemble closely the Frenkel-Kontorova
model, at least for $d\gg\sigma$, where the lowest Fourier component at
period $l$ would dominate the effective corrugation experienced by the
chain atoms.

%------------------------------------------------------------------------
\begin{figure}
\centering
\includegraphics*[width=0.5\textwidth,angle=0]{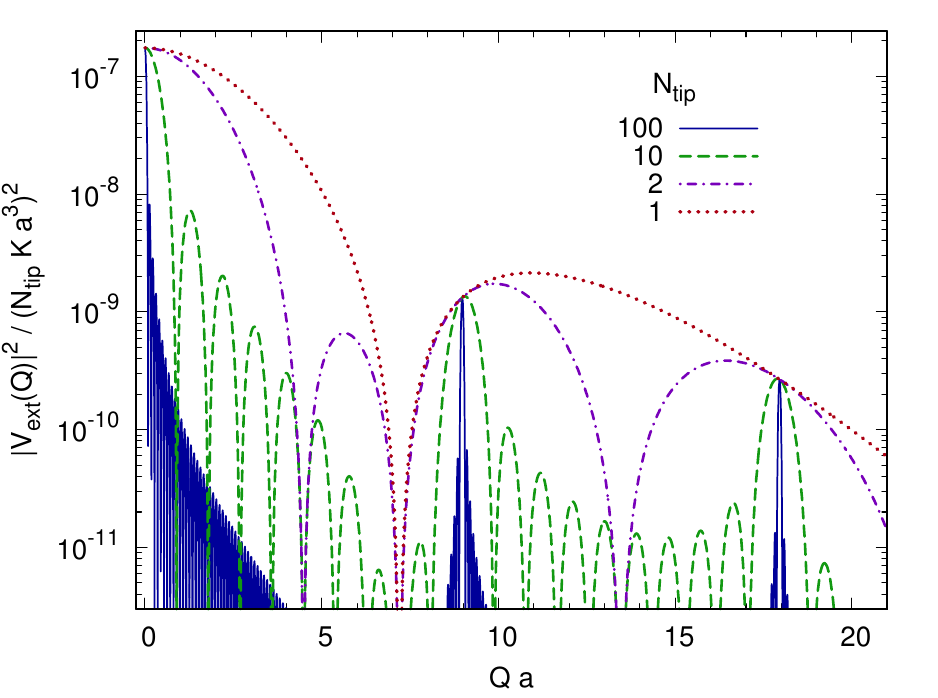}
\caption{\label{fig:tips_fft}
  Comparison of the Fourier transforms $|V_{\rm ext}(Q)|^2$ of the
  slider-chain interaction, for simulated sliders consisting in the sum
  of $N_{\rm tip}$ LJ contributions placed regularly at an interval
  $l=0.7\, a$.
  The other parameters for $V_{\rm ext}$ are the standard ones used in
  this work: $\sigma=0.5 \, a$, $d = 0.465 \, a$, $V_0 = 5\times
  10^{-4}\, Ka^2$.
}
\end{figure}
%------------------------------------------------------------------------

Here we focus on a finite-size slider, representative of a microscopic
contact such as an AFM tip.
As an example, Fig.~\ref{fig:tips} compares the dynamic friction as a
function of speed for $N_{\rm tip}=1, 2, 10, 100$, in the usual
conditions $\sigma=0.5 \, a$, $d=0.92 \, \sigma$, and $\gamma=0.1 \,
(m/K)^{1/2}$.
The resulting friction curves exhibit an additional and interesting
dependence upon the tip size.
The initial simplicity for $N_{\rm tip}=1$ is lost with the addition of
oscillations for 2 and 10, until simplicity is again recovered when we
reach $N_{\rm tip}=100$.
These features can be understood by examining the effect of increasing the
number of LJ particles on the Fourier transform of the potential.
As illustrated in Fig.~\ref{fig:tips_fft}, as the number of tip atoms
grows, $|V_{\rm ext}(Q)|^2$ exhibits an increasing number of
oscillations, with $Q$-space period $2\pi / [(N_{\rm tip}-1)\, l]$,
generated by the sharp edges of a slider whose overall size is $(N_{\rm
  tip}-1)\,l$.

These fast oscillations produce sets of wave vectors $\{ Q^0_{i} \}$
such that $V_{\rm ext}(Q^0_{i})=0$: these wavevectors yield friction
dips for the velocities $v_{\SL}$ matching the condition $Q^0_{i}
v_{\SL} =\omega(Q^0_{i})$.
These dips are especially visible for $N_{\rm tip}=10$, dashed line in
Fig.~\ref{fig:tips}.
As $N_{\rm tip}$ grows, the distance between the zeroes of $V_{\rm ext}(q)$
decreases, the number of friction dips increases, and their visibility
decreases: beyond a certain slider size the wave vector $2\pi/[(N_{\rm tip}-1)\, l]$ 
of this oscillation becomes shorter than the phonon broadening,
of the order of $\gamma/v_{\rm s}$, introduced by the damping term.

In addition to the size-related oscillations, for increasing $N_{\rm tip}$, 
$|V_{\rm ext}(Q)|^2$ develops stronger and sharper Bragg peaks
related to the periodicity of the atoms in the tip.
The relevant $G$-vectors are located at integer multiples of $G_1 = 2\pi/l
\simeq 8.976/a$, especially evident in the solid curve of Fig.~\ref{fig:tips_fft}.
The resonant strengths $|V_{\rm ext}(G_i)|^2$ grow with $N_{\rm tip}^2$
--- note that in Fig.~\ref{fig:tips_fft} $|V_{\rm ext}(Q)|^2$ is
reported divided by $N_{\rm tip}^2$, so that the Bragg peaks retain the
same heights for all curves.
These intensities are modulated by the ``atomic form factor''
represented by $|V_{\rm ext}(Q)|^2$ for $N_{\rm tip}=1$, namely the
dotted curve of Fig.~\ref{fig:tips_fft}.
For the selected spacing $l=0.7\,a$, the near coincidences of $G_1\simeq
Q_2$ and $G_3 \simeq Q_5$, among the resonant vectors $Q_j$ reported in
Table~\ref{tbl:sol}, imply that the corresponding resonant peaks at
$v_{{\SL}\,2}\simeq 0.22 \, v_{\rm s}$ and $v_{{\SL}\,5}\simeq
0.07 \, v_{\rm s}$ are especially prominent in the $F$ profile for
$N_{\rm tip}=10$ and $100$ -- dashed and solid curves of
Fig.~\ref{fig:tips}.
As the spacing $l$ of the slider atoms can be controlled independently
of other model parameters, the analytic formulation of the present work
provides a way to engineer a specially crafted polyatomic slider with
Bragg peaks suitably placed in order to enhance or suppress specific
resonances, thus tuning the speed dependence of friction practically at
will.

%----------------------------------------------------------------------------
\section{Conclusions}\label{conclusions}
%----------------------------------------------------------------------------

Based on an exact LRT starting point and brought to a viable formula by
means of well-understood approximations, we presented an analytic result
for the dynamic sliding friction for a minimal model for sliding in the
weak-interaction regime.
This formula provides a transparent physical decomposition of the friction
force into products of equilibrium dynamical properties of the unperturbed
chain and mechanical properties of the slider-chain interaction.
Within this approximation, the sliding of a small rigid slider at a
certain velocity excites selectively those phonons in the substrate
moving at the same phase velocity.
Specific resonances occur when these phonons share both the {\em same
  phase velocity and group velocity} as that of the slider.

Simple as this approach is, it shows two strong features.
The first is that it is conceptually and practically straightforward to calculate friction.
The second is that at least in the test system adopted it reproduces with
striking accuracy the friction observed in much more laborious numerical
simulations,\cite{Apostoli17} whose output serves as a validating
numerical experiment.

Importantly, the analytic relation obtained provides an explicit,
therefore powerful insight into weak, smooth, dry sliding friction.
The explicit analytic form potentially allows an easy tuning of the
properties of the sliders and their interactions with the purpose of
tailoring a desired speed and load dependence of dynamical friction.

We purposely conducted this study for a simple idealized 1D model, as
opposed to a more specific one, because of the clarity with which the
results and phenomena could be uncovered as a function of parameters.
The understanding and the stunning accuracy obtained suggest that this
study may serve as a guide for future applications of this method to
more realistic 3D systems.
Suitable approximations to the structure factor $S_{nn}$ of 3D
substrates, and of slider-substrate interaction, could be used for
approximate predictions of friction for realistic interfaces, at least
in the weak-interaction regime, at arbitrary speed and variable load,
including regimes such as low speeds which simulations cannot reach.
We should reiterate here that by its linear-response basis the method only
applies to smooth sliding at finite speed, while it fails to predict
anything about the transition from dynamic to static friction, and thus
about intrinsically nonlinear regimes such as stick-slip.
This approach on the other hand may become quite valuable in providing
analytical clues to the behavior of friction including systems and
circumstances where its applicability might be considered borderline.

\section*{Acknowledgment}
Work in Trieste was carried out under ERC Grant 320796 MODPHYSFRICT. The COST
Action MP1303 is also gratefully acknowledged.

%++++++++++++++
\appendix
%++++++++++++++

\section{Derivation of the LRT expression in Eq.~\eqref{eq:finalW}} \label{chiandF:app}

To obtain Eq.~\eqref{eq:finalW}, we start from Eq.~\eqref{Et_LJ:eqn} and
represent $\chi^R_{nn}$ as a Fourier transform.
To enforce the crystalline translational invariance \cite{Strinati88} we
write:
\begin{widetext}
\begin{eqnarray} \label{Xnn_F:eqn}
  \chi_{nn}^R(x,x';t-t') = \sum_{G} \int_{-\infty}^{+\infty} 
  \frac{\ud Q}{2\pi} \int_{-\infty}^{+\infty} \frac{\ud\omega}{2\pi} \;
  \nep^{-i\omega(t-t')} \; \nep^{iQx} \;
  \chi_{nn}^R(Q,Q+G;\omega) \; \nep^{-i(Q+G)x'}
  \,,
\end{eqnarray}
\end{widetext}
where the sum over $G$ runs over the reciprocal lattice vectors $G=2\pi
n/a$.
Since the only time dependence of the perturbing potential is a space-shift, $V_{\rm ext}(x, t)\,=\,V_{\rm ext}(x-v_{\SL} t, 0)$, its standard Fourier representation is:
\begin{eqnarray} \label{vex_F:eqn}
  V_{\rm ext}(x,t) =
  \int_{-\infty}^{+\infty} \frac{\ud q}{2\pi} \,
  \nep^{iq(x-v_{\SL} t)}\, V_{\rm ext}(q)  \;. 
\end{eqnarray}
Notice that $V_{\rm ext}(-q)=V_{\rm ext}^*(q)$ since $V_{\rm ext}(x-v_{\SL}t, 0)$ 
is a real function.
To calculate the average dissipation we average $\dot{E}=W$ in 
Eq.~\eqref{Et_LJ:eqn} over a period of time $\tau = a/v_{\SL}$.
Inserting all Fourier transforms we have:
\begin{widetext}
\begin{eqnarray*}
  F(v_{\SL}) = \frac{1}{v_{\SL}} \overline{W} 
  &=& -\frac{1}{v_{\SL} \tau} \, \int_{0}^{\tau} \!\ud t
  \int_{-\infty}^{+\infty} \!\ud x \int_{-\infty}^{+\infty} \!\ud x'
  \int_{-\infty}^{+\infty} \! \ud t' 
  \int_{-\infty}^{+\infty} \!\frac{\ud q}{2\pi} \,
  \int_{-\infty}^{+\infty} \! \frac{\ud q'}{2\pi} 
  \sum_{G}  \int_{-\infty}^{+\infty}  \!\frac{\ud Q}{2\pi}
  \int_{-\infty}^{+\infty}\! \frac{\ud \omega}{2\pi}
  \times \nonumber \\ && \phantom{-}
  \nep^{i(q+Q)x} \nep^{-i(q'+Q+G)x'}
  \nep^{-it(\omega + qv_{\SL})} \nep^{it'(\omega+q' v_{\SL})}
  (-i\omega)\chi_{nn}^R(Q,Q+G;\omega) V_{\rm ext}(q)V_{\rm ext}^*(q')
\,.
\end{eqnarray*}
Integration over $x$ and $x'$ yields two Dirac-delta distributions for
$q$ and $q'$, leading to:
\begin{eqnarray}
  F(v_{\SL}) &=& -\frac{1}{v_{\SL} \tau}
  \int_{0}^{\tau} \!\ud t \int_{-\infty}^{+\infty} \!\ud t'
   \sum_{G} \int_{-\infty}^{+\infty} \frac{\ud Q}{2\pi}
  \int_{-\infty}^{+\infty} \frac{\ud\omega}{2\pi}  
  \times \nonumber \\ && \phantom{-}
  \nep^{-it(\omega -Q v_{\SL})}
  \nep^{it'(\omega-(Q+G)v_{\SL})}
  (-i\omega)\,\chi_{nn}^R(Q,Q+G;\omega) V_{\rm ext}^*(Q)V_{\rm ext}(Q+G)
  \,.
\end{eqnarray}
The integral over $t'$ gives a Dirac delta $2\pi
\delta(\omega-(Q+G)v_{\SL})$:
\begin{eqnarray*}
  F(v_{\SL})
  = -\frac{1}{v_{\SL} \tau}
  \sum_{G} \int_{0}^{\tau} \!\ud t \, \nep^{-it G v_{\SL}} \,
  \int_{-\infty}^{+\infty} \frac{\ud Q}{2\pi}
  (-i(Q+G)v_{\SL}) \,
  \chi_{nn}^R(Q,Q+G;(Q+G) v_{\SL}) V_{\rm ext}^*(Q)V_{\rm ext}(Q+G)
\,.
\end{eqnarray*}
\end{widetext}
The $t$ integral over a period $\tau$ now involves
\begin{equation} \nonumber
\frac{1}{\tau} \int_{0}^{\tau} \!\ud t \, \nep^{-it G v_{\SL}} =
\delta_{G,0} \;,
\end{equation}
due to $\tau=a/v_{\SL}$.
Hence:
\begin{eqnarray}\label{eq:simpl1}
  F(v_{\SL}) &=& -\frac{1}{v_{\SL}} \int_{-\infty}^{+\infty} \frac{\ud Q}{2\pi} \,
  (-iQv_{\SL})\,
  \times \nonumber \\ && \phantom{-}
  \chi_{nn}^R(Q,Q;Q v_{\SL}) \, |V_{\rm ext}(Q)|^2 \,.
\end{eqnarray}
Equation~\eqref{eq:simpl1} shows that the relevant frequency $\omega$
contributing is related to $Q$ via
\begin{equation}
  \widetilde{\omega}_Q=Qv_{\SL} \,. 
\end{equation}

Next, we use the standard properties of $\chi_{nn}^R(Q,Q;\omega)$ to
show that only the {\em imaginary part} of $\chi_{nn}^R$ contributes.
Indeed, hermiticity of the density operator implies that
$\chi_{nn}^R(Q,Q;\omega)^* = \chi_{nn}^R(Q,Q;-\omega)$ and
$\chi_{nn}^R(Q,Q;\omega) = \chi_{nn}^R(-Q,-Q;\omega)$.
Since $|V_{\rm ext}(Q)|^2$ is an even function of $Q$, only the even
part of $Q \chi_{nn}^R(Q,Q;Q v_{\SL})$ contributes to the
integral.
Simple algebra shows that
\begin{equation}
  F(v_{\SL}) = -\frac{2}{v_{\SL}} \int_{0}^{+\infty} \frac{\ud Q}{2\pi} \,
  \widetilde{\omega}_Q \,
  \mathrm{Im} \chi_{nn}^R(Q,Q;\widetilde{\omega}_Q)
  \, |V_{\rm ext}(Q)|^2 \,, 
\end{equation}
which is the first form of Eq.~\eqref{eq:finalW}.

The second form comes from the use of the fluctuation-dissipation
relation\cite{Vignale05}
\begin{equation}
  \mathrm{Im} \chi_{nn}^R(Q,Q;\omega) =
  - \frac{1}{2}\left( 1 - \nep^{-\beta\hbar\omega} \right)
  S_{nn}(Q,Q;\omega)
  \,,
\end{equation}
where $S_{nn}(Q,Q;\omega)$ is the structure factor:
\begin{widetext}
\begin{eqnarray} \label{eqn:S_nn}
S_{nn}(Q,Q;\omega) &=& 
\lim_{V\to \infty} \frac{1}{V} \int_{V} \! \ud x' \int_{V} \! \ud x
\; \nep^{-iQ(x-x')}  \; 
\frac{1}{\hbar} \int_{-\infty}^{+\infty} \! \ud t \;
\nep^{i\omega t} \; \langle \hat{n}(x,t) \hat{n}(x',0) \rangle 
\nonumber \\ &=&
\frac{1}{\hbar} \lim_{V\to \infty} \frac{1}{V}
\int_{-\infty}^{+\infty} \! \ud t \; \nep^{i\omega t} \; 
\langle \hat{n}_Q(t) \hat{n}_{-Q}(0) \rangle \;.
\end{eqnarray}
\end{widetext}
Here
\begin{equation}
\hat{n}_Q(t) = \int_{V} \ud x \; \nep^{-iQx} \; \hat{n}(x,t) 
\end{equation}
is the Fourier transform of the density operator (in Heisenberg
representation) for a system in a 1D ``volume'' $V$ with
periodic boundary conditions (PBC).

\section{Evaluation of the structure factor for the harmonic chain} \label{structure:sec}

In this appendix we calculate the structure factor for a harmonic chain.
We use the standard \cite{AshcroftAppN} displacement $\hat{u}_j(t)$ from the
equilibrium position $x^{(0)}_{j}=aj$ for each atom $j$ to write the
Fourier transform of the density operator as:
\begin{equation}
  \hat{n}_Q(t) = \sum_{j=1}^N \nep^{-iQaj} \; \nep^{-iQa\hat{u}_j(t) }
  \,,
\end{equation}
where $N$ is the number of atoms, hence $V=Na$ is the volume, and PBC
are assumed.
Using this expression and the lattice translational invariance of the
problem we can write the structure factor of Eq.~\eqref{eqn:S_nn} as:
\begin{eqnarray} 
  S_{nn}(Q,Q;\omega)
  &=&
  \lim_{N\to \infty} 
  \frac{1}{\hbar a} \int_{-\infty}^{+\infty} \! \ud t \; \nep^{i \omega t}
  \times \nonumber \\ &&
  \sum_{j=1}^N \nep^{-iQaj}
  \mean{\nep^{-iQ\hat{u}_{j}(t)} \nep^{+iQ\hat{u}_0(0)} }
  \,.\,
\end{eqnarray}
To evaluate the averages, we use the known Gaussian identity
$\mean{\nep^{\hat{A}} \nep^{\hat{B}} } =
\nep^{ \frac{1}{2}\mean{\hat{A}^2}
  + \frac{1}{2} \mean{\hat{B}^2} + \mean{\hat{A}\hat{B}} }$
valid for harmonic-oscillator operators.\cite{mermin1966short}
A direct application of this formula to our case of interest, together
with translational invariance in time and lattice cell index, leads to:
\begin{eqnarray}\label{prodexp:eq}
  \mean{\nep^{-iQ\hat{u}_{j}(t)}\nep^{iQ\hat{u}_{0}(0)}} =
  \nep^{-Q^2\mean{\hat{u}_{j}(t) \hat{u}_{j}(t)}}\,
  \nep^{Q^2\mean{\hat{u}_{j}(t)\hat{u}_{0}(0)}}
  \,.
\end{eqnarray}
By expressing the displacement operators $\hat{u}_j$ in terms of
standard harmonic oscillators for the phonons\cite{AshcroftAppN}
\begin{eqnarray}\label{hjt:eq}
  \hat{u}_{j}(t)
  &=&
%  \nep^{iH_{\rm harm} t/\hbar} \hat{u}_j \nep^{-i H_{\rm harm} t/\hbar} = 
  \frac{1}{\sqrt{N}} \sum_{k\neq 0}^{\rm BZ}
  \nep^{ikaj} \sqrt{\frac{\hbar}{2m\omega(k)}} 
  \times \nonumber \\ &&
  \left( \nep^{-i\omega(k) t}\, \opb{k} +
  \nep^{i\omega(k) t} \, \opbdag{-k} \right)
  ,
\end{eqnarray}
we can easily calculate:
\begin{widetext}
\begin{eqnarray} \label{term2:eq}
 \mean{\hat{u}_{j}(t)\hat{u}_{j}(t)} &=& \frac{1}{N} \sum_{k}^{\BZzero}
  \frac{\hbar}{2 m \omega(k) } \left[ 2n_B(k) + 1 \right] \nonumber \\
  \mean{\hat{u}_{j}(t)\hat{u}_{0}(0)} &=&
  \frac{1}{N} \sum_{k}^{\BZzero} \frac{\hbar}{2 m \omega(k) }
  \left[n_{\rm B}(k)\nep^{i \omega(k) t} +
    (n_{\rm B}(k)+1)\, \nep^{-i\omega(k) t} \right]  \nep^{ikaj}
  \,,
\end{eqnarray}
where $n_B(k)=1/(\nep^{\beta\hbar\omega(k)}-1)$ is the Bose distribution
factor.
By combining Eq.~\eqref{term2:eq} with Eq.~\eqref{prodexp:eq}, we
finally arrive at the following expression for the structure factor of
an harmonic chain (in the thermodynamic limit):
\begin{equation} \label{eq:snn}
\left\{
\begin{array}{rcl}
  S_{nn}(Q,Q;\omega)
  &=&  \displaystyle
  \frac{1}{\hbar\,a}
  \int_{-\infty}^{+\infty}\!\! \ud t \; \nep^{i \omega t}
  \sum_{j=-\infty}^{+\infty}  \nep^{-iQaj} \, \nep^{-Q^2 \, \Phi_j(t,\beta)} \\
  \Phi_j(t,\beta)
  &=& \displaystyle
  a \int_{-\frac{\pi}{a}}^{+\frac{\pi}{a}}  \frac{\ud k}{2\pi} \,
  \frac{\hbar}{2 m \omega(k) } \left\lbrace [2 n_{\rm B}(k)+1] -
  \left[n_{\rm B}(k)\nep^{i \omega(k) t} +
    (n_{\rm B}(k)+1)\, \nep^{-i\omega(k) t} \right]
  \nep^{ikaj} \right\rbrace 
\end{array}
\right.
.
\end{equation}
\end{widetext}
We could find no way to evaluate this exact expression.
To proceed we resort to the standard one-phonon \cite{AshcroftAppN}
expansion of the exponential
\begin{equation}\label{linearexp:eq}
  \nep^{-Q^2 \Phi_j(t,\beta)} \simeq 1 - Q^2 \Phi_j(t, \beta)  \,.
\end{equation}
Such an approximation is rather drastic, especially in low dimension:
indeed, the argument $Q^2 \, \Phi_j(t,\beta)$ of the exponential in
Eq.~\eqref{linearexp:eq} is not always small.
One can argue that the $Q$ dependence can be regularized by a
sufficiently fast decay of the potential $|V_{\rm ext}(Q)|^2$, thus
legitimating an expansion.
Nevertheless, even if $Q$ is assumed to be small one can show that in
1D $\Phi_j(t,\beta)$ would diverge linearly in time (or
logarithmically, for $T = 0$) for large $t$.
The divergence of $\Phi_j(t,\beta)$ leads to a factor $\exp\left(-Q^2
\Phi_j(t,\beta)\right)$ which drops to zero for large $t$ --- hence the
absence of {\em elastic Bragg peaks}, proportional to $\delta(\omega)$,
in 1D --- while the linearized one-phonon expression actually
diverges.
On one hand, such subtleties are just an artefact of the 1D
toy problem we have considered, and should not influence applications of
our theory to more realistic situations.
On the other hand, we find that even for a strictly 1D toy
problem, the (inevitable) presence of dissipation provides a cure for
the problem.
Indeed the decaying exponential term in Eq.\eqref{eq:Slangevin} of Sect.~\ref{sec:evaluation}, 
introduced to explicitly take into account dissipation,
kills the divergences of $\Phi_j(t,\beta)$ at large $t$ for any temperature $T$.

Consider now the resulting expression for the one-phonon {\em inelastic}
structure factor:
\begin{widetext}
\begin{equation}
S_{nn}^{\oneph}(Q,Q;\omega>0) =Q^2 \int_{-\infty}^{+\infty}\!\! \ud t 
 \int_{-\frac{\pi}{a}}^{+\frac{\pi}{a}}  \frac{\ud k}{2\pi} \, \frac{n_{\rm B}(k)+1}{2 m \omega(k)}
  \, \nep^{i(\omega-\omega(k)) t} \sum_{j=-\infty}^{+\infty}   \nep^{i(k-Q)aj}
\,.
\end{equation}
\end{widetext}
Here we have applied the approximation \eqref{linearexp:eq} to
Eq.~\eqref{eq:snn}, and kept only the term that survives for $\omega>0$.
We dropped all the terms inside the integrand which do not depend on
time, as they give a vanishing contribution in 1D, and would
lead to elastic $\delta(\omega)$ contributions in dimensions $D>2$; we
also dropped ``counter-rotating'' terms which oscillate as
$\nep^{i(\omega+\omega(k)) t}$.
Now we make use of the {\em periodic delta function}, or {\em Dirac
  comb}, identity:
\begin{equation}\label{sumj}
  \sum_{j=-\infty}^{\infty} \nep^{i (k -Q) a j} = \frac{2\pi}{a} \sum_{G} \delta(k-Q-G) \;,
\end{equation}
where $G=\frac{2\pi n}{a}$ and $n$ is any integer.
Hence, combining the integral over $k$ on the first BZ with the
reciprocal-lattice $G$ summation resulting from Eq.~\eqref{sumj}, one
obtains an unrestricted integral on a variable $k'=k-G$ spanning the
extended BZ scheme, i.e.\ the entire $-\infty<k'<+\infty$ range.
The resulting $\delta(k'-Q)$ identifies $k'$ with $Q$.
Eventually, performing the integral over $t$, which yields a
delta-function in $\omega$, we obtain:
\begin{equation} \label{eq:dyn_S_1ph}
  S_{nn}^{\oneph}(Q,Q;\omega>0) = \frac{\pi Q^2 }{m a \omega(Q)}   
  \left[n_{\rm B}(Q)+1\right]
  \delta(\omega-\omega(Q))
\,,
\end{equation}
which is equivalent to Eq.~\eqref{eq:dyn_S_1ph_main}.
Note that in this expression the dispersion $\omega(Q)$ is intended in
the extended BZ scheme.

%------------------------------------------------------------------
%  BIBLIOGRAPHY
%------------------------------------------------------------------
 
\bibliography{biblio}

\end{document}